\documentclass[twocolumn,showpacs,preprintnumbers,amsmath,amssymb]{revtex4}
\usepackage{epsfig}

\newcommand{\beq}{\begin{equation}}
\newcommand{\eeq}{\end{equation}}
\newcommand{\bes}{\begin{subequations}}
\newcommand{\ees}{\end{subequations}}
\newcommand{\bea}{\begin{eqnarray}}
\newcommand{\eea}{\end{eqnarray}}
\newcommand{\ba}{\begin{array}}
\newcommand{\ea}{\end{array}}
\newcommand{\beqn}{\begin{eqnarray*}}
\newcommand{\eeqn}{\end{eqnarray*}}

\newcommand{\f}[2]{\frac{#1}{#2}}
\newcommand{\g}{\gamma}

\newcommand{\om}{\omega}

\def\nn{\nonumber}

\begin{document}

\title{Crossover from ballistic to diffusive thermal transport in quantum Langevin dynamics study of a harmonic chain connected to self-consistent reservoirs}
\author{Dibyendu Roy}
\affiliation{$^1$Raman Research Institute, Bangalore  560080, India}
\date{\today} 

\begin{abstract}
 Through an exact analysis using quantum Langevin dynamics, we demonstrate the crossover from ballistic to diffusive thermal transport in a harmonic chain with each site connected to Ohmic heat reservoirs. The temperatures of the two heat baths at the boundaries are specified from before whereas the temperatures of the interior heat reservoirs are determined self-consistently by demanding that in the steady state, on average, there is no heat current between any such (self-consistent) reservoir and the harmonic chain. Essence of our study is that the effective mean free path separating the ballistic regime of transport from the diffusive one emerges naturally. 
\end{abstract}      

\maketitle

Fourier's law is an old empirical law stating connection between heat current density and spatially varying temperature field. But it is still not clear what are the necessary and sufficient conditions for the validity of Fourier's law of heat transport \cite{bonetto00, lepri01, wu07}. Heat conduction through a one-dimensional ordered harmonic chain connected with two heat reservoirs at different temperatures shows ballistic nature. Also it is well established that heat transport in one-dimensional momentum conserving systems (absence of external potentials) does not follow Fourier's law \cite{dhar01, narayan02}. Heat conduction in a long harmonic chain connected to self-consistent reservoirs at every site shows diffusive behaviour qualifying system size independent thermal conductivity \cite{Bolsterli70, bonetto04, AbhiDib06}. The transition from ballistic to diffusive dynamics in thermal and electrical transport has recently received a lot of attention. In a recent letter, Wang \cite{wang07} has reported to have obtained quantum thermal transport from classical molecular dynamics using a generalised Langevin equation of motion. Based on a ``quasiclassical approximation'', the author claims to reconcile the quantum ballistic nature of thermal transport with diffusive one in a one-dimensional quartic on-site potential model. In Ref. \cite{steinigeweg07} the authors have studied the transition from diffusive to ballistic dynamics for a class of finite quantum models by an application of the time-convolutionless projection operator technique. Here through an exact analysis using quantum Langevin dynamics, we demonstrate the crossover from ballistic to diffusive thermal transport in a harmonic chain connected to self-consistent reservoirs.

Consider heat conduction through a one-dimensional ordered harmonic chain of particles $l=1,2...N$ with unit masses which are connected by harmonic springs of equal strengths. The Hamiltonian of the system is 
\bea
H=\sum_{l=1}^{N}\f{\dot{x_l}^2}{2}+\sum_{l=0}^{N}\f{(x_{l+1}-x_l)^2}{2},
\eea 
where $\{x_l\}$ are Heisenberg operators, correspond to particle displacements about some equilibrium configuration. We choose the boundary conditions $x_0=x_{N+1}=0$. All the particles are connected to Ohmic heat reservoirs with coupling strength controlled by dissipation constant $\g_l$. We set $\g_l=\g$ for $l=1,N$
and $\g_l=\g'$ for $l=2,3..N-1$. This allows us to tune the coupling $(\g')$ between self-consistent reservoirs and the chain sites without affecting the couplings at the end reservoirs. The temperatures of the first and last reservoirs are fixed as $T_1=T_L$ and $T_N=T_R$ respectively. The temperatures of the attached interior reservoirs $\{T_l\}$ are determined self-consistently by the condition of net zero heat current from the side reservoir to the chain. Slightly different version of this model has been studied in \cite{bonetto04, AbhiDib06} for infinitely long chain length. The time-evolution of the chain particles is governed by a combination of Hamiltonian and stochastic dynamics, where the nearest neighbour harmonic potentials form the Hamiltonic part and the stochastic influence  comes from the connection of each chain site with Ohmic heat bath.  As pointed out in \cite{bonetto04}, the influences of the interior reservoirs can be considered  as the effect of degrees of freedom not present in the Hamiltonian. It is also shown in \cite{bonetto04} that in the limit $N\to\infty$, the steady state is a local equilibrium state with a temperature profile satisfying Fourier's law with a temperature independent, finite thermal conductivity for the classical model. The quantum version of this model is studied in \cite{AbhiDib06} where a finite, temperature dependent thermal conductivity is found in the quantum regime.    

\begin{figure}[t]
\begin{center}
\includegraphics[width=7.5cm]{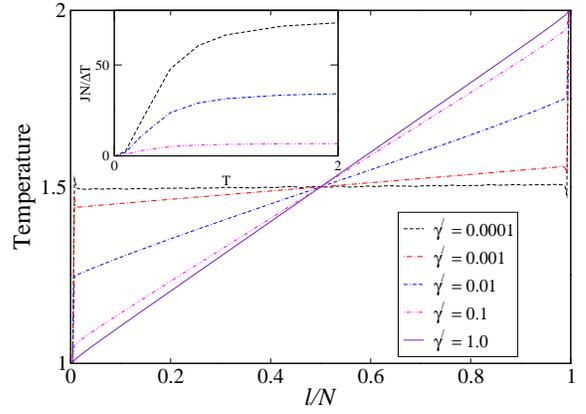}
\end{center}
\caption{ Plot of the temperature profile $\{T_l\}$ as a function of scaled length $l/N$ for different $\g'$ with $\g=1.0$ and $N=256$. The inset shows temperature dependence of scaled current for different $\g'$ with above values of $\g, N$. Here mean free path $\ell=3/\g'$. }
\label{cross2}
\end{figure}

The quantum Langevin equations of the chain sites are, 
\bea
\ddot{x_l}=-2x_l + x_{l-1} + x_{l+1} -\g_l\dot{x_l} + \eta_l~~~{\rm for}~~l=1,2,3...N,\nn
\eea
where $\eta_l$ is the noise generating from the $l$th reservoir. The correlations of noises, are such that the distributions of normal modes in isolated reservoirs follow Bose-Einstein statistics. Noise-noise correlation in the frequency domain is given by,
\bea
\f{1}{2}\langle\eta_l(\om)\eta_m(\om')+\eta_l(\om')\eta_m(\om)\rangle=\f{\g\om}{2\pi}\coth(\f{\om}{2T_l})\delta(\om+\om')\delta_{lm},\nn
\eea
with $\hbar=K_B=1$. 
\begin{figure}[t]
\begin{center}
\includegraphics[width=8cm]{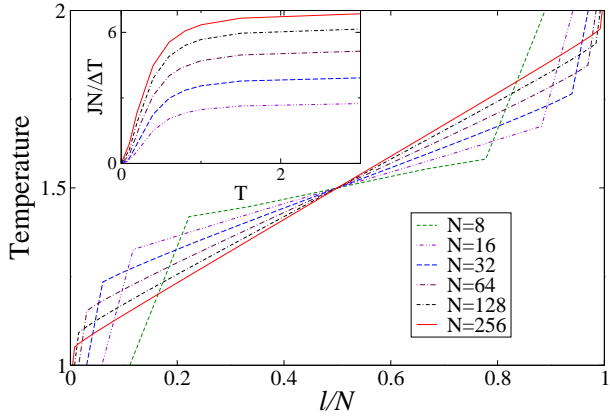}
\end{center}
\caption{ Plot of the temperature profile $\{T_l\}$ as a function of scaled length $l/N$ for different $N$ with $\g=1.0$ and $\g'=0.1$. The inset shows temperature dependence of scaled current for different $N$ with above values of $\g,\g'$. Here mean free path $\ell=30$. }
\label{cross1}
\end{figure}
Now our first task is to determine the temperature profile of the interior reservoirs from the self-consistent condition. For this we write down the heat current from the reservoir to the chain and then expand it in the linear response regime; the current from the $l$th reservoir to the chain is given as \cite{AbhiDib06}
\bea
J_l=\sum_{m=1}^{N}\f{\g_l\g_m}{\pi}\int_{-\infty}^{\infty}d\om \f{\om^4}{4T^2}{\rm cosech}^2(\f{\om}{2T})|G_{lm}|^2\f{(T_l-T_m)}{\pi},\nn\\
\label{cur1}
\eea
where $G(\om)$ is inverse of a tridiagonal matrix with offdiagonal elements equal to $-1$ and diagonal elements equal to $2-\om^2-i\g'\om$ except the ends which are equal to $2-\om^2-i\g \om$. Also $T=(T_L+T_R)/2$. We set $J_l=0$ for $l=2,3...N-1$ and solve $N-2$ linear equations numerically to find $\{T_l\}$ profile. In Fig.\ref{cross1}  we plot $\{T_l\}$ for different lengths of the chain for some fixed small value of $\g'$. In the limit $\g'<<1$ we find the temperature profile is scaled as 
\bea
T_1&=&T_L~,~T_N=T_R~{\rm and}\nn\\
T_l&=&T_L+\delta+\f{2\delta}{\ell}(l-2)~~{\rm for}~~l=2,3...N-1,\nn\\
{\rm with}~~ \delta&=&\f{\Delta T}{2(1+N/\ell)}~,
\eea
where $\ell=3/\g'$ and $\Delta T=T_R-T_L$. Here $\delta$ is the jump in the temperature at the boundaries. The above scaling relation can be derived from a persistent random walk model of phonons in analogy with one for electrons \cite{DibAbhi07}; here $\ell$ is interpreted as the mean free path of the phonons. We also plot $\{T_l\}$ profile in Fig.\ref{cross2} for fixed length $N=256$ with different values of $\g'$. When $\g'$ tends to zero (then $\ell$ goes to infinity), the heat transport in the chain is ballistic (becomes completely ballistic at $\g'=0$), and the temperature profile is flat as shown in Fig.\ref{cross2}. With increasing $\g'$ system transits through a mixed transport regime towards a diffusive one for sufficiently large value of $\g'$ where $\ell$ is much smaller than the system size; then $\{T_l\}$ profile is linear. For larger value of $\g'$, the temperature profile becomes linear for smaller system sizes.

To find the heat current through the wire from the left to the right heat baths we can use the heat current expression of Eq.(\ref{cur1}) with $l=1$ or $N$. But we notice that Eq.(\ref{cur1}) then requires the temperatures at the boundaries accurately as these terms contribute significantly. So if we want to use the scaling form of the temperature which is not so good at the boundaries for smaller wire sizes, it is better to evalute current in the middle bond of the wire. Current through $(l,l+1)$ spring of the chain is given by \cite{AbhiDib06}
\bea
&&J_{l,l+1}=\langle x_l \dot{x}_{l+1}\rangle \nn \\
&=&-\sum_{m=1}^N\f{\g_m T_m}{\pi}\int_{-\infty}^{\infty}d\om \f{\om^3}{4T^2}{\rm cosech}^2(\f{\om}{2T})Im[G_{lm}G^\ast_{l+1m}]\nn\\
\eea
Now using the numerical solution for $\{T_l\}$ we first evalute the heat current with varying temperature for different $N$ and $\g'$, and plot it in the inset of Figs.\ref{cross1} and \ref{cross2} respectively. $J_{l,l+1}$ (call $J$) is independent of $l$ and we calculate it in the bulk for accuracy. Using the scaling form of $\{T_l\}$ we find
\bea
J=\f{\kappa(T)~\Delta T}{(N+\ell)}
\eea
$\kappa(T)$ is the temperature dependent thermal conductivity of the infinite chain. Above current expression is exact for larger size of the chain but for smaller size there will be correction from the boundaries. It clearly shows that for $N>>\ell$ transport is diffusive satisfying Fourier's law and in the opposite limit the current is independent of $N$ (ballistic). We clarify that the cross-over from ballistic to diffusive behaviour in transport depends on the effective length scale of the problem and can be controlled here by tuning $\ell$, i.e., $\g'$. This model has similarity to the one-dimensional quartic onsite potential model \cite{wang07} if one identifies $\g'$ with the strength of the quartic onsite potential. But in the quartic onsite potential model the temperature and the strength of quartic potential are conjugate to each other, i.e., for a fixed strength of the quartic potential, increasing the temperature one can cross-over from ballistic to diffusive regime of transport; similarly for a constant temperature, changing the strength of quartic potential one can tune from ballistic to diffusive transport. But in case of the self-consistent reservoir model the temperature and the strength of the coupling to the reservoirs $(\g')$ are independent parameters, not affecting each other.

In conclusion we have demonstrated both ballistic and diffusive regime of thermal transport within a single analysis of quantum Langevin dynamics. This is contrary to the remark made by author in Ref.\cite{wang07}. As discussed nicely in \cite{bonetto00, Bolsterli70}, it is a big challenge to derive Fourier's law for a system from microscopic Hamiltonian bulk dynamics. One can think of the problem as either (a) the system in microcanonical ensemble evolving towards equilibrium from an initial arbitrary distribution and study the relaxation mechanism from different correlations like energy-energy; or (b) the system is kept in a non-equlibrium steady state by connecting it at the boundaries with stochastic or mechanical reservoirs and then determine the size dependence of the steady state current . From a large number of numerical and a few analytical studies \cite{bonetto00, lepri01} it is believed that the chaotic behaviour generating from nonintegrability is essential criterion for realising Fourier's law in classical systems. Analogously from the numerical study of quantum systems with the coupling to external heat baths \cite{Saito96, Monasterio05} or without \cite{Steinigeweg06}, it is argued that the emergence of diffusive behaviour is related to onset of quantum chaos.  Now we try to analyse the underlying mechanism of getting diffusive behaviour in this self-consistent reservoir model. Originally Bolsterli, Rich and Visscher \cite{Bolsterli70} proposed the self-consistent reservoir model to incorporate phenomenologically the interactions of phonons with other degrees of freedom such as electron's charge and spin present in the physical system. Due to stochastic interactions with the internal reservoirs the inherently non-ergodic harmonic chain becomes ergodic. Here the self-consistent reservoirs provide the mechaism of scattering for phonons which is very much essential to get diffusive behaviour. It can also be posed in a different way that the self-consistent reservoirs act as the environment in a persistent random walk of phonon in a lane and break down the coherent nature of transport. In this context this model is similar to the models of particle transport studied in \cite{DibAbhi07} again with self-consistent particle reservoirs and in \cite{Esposito05} with heat baths modelling the dissipative environment.  Now we point out certain inconsistencies in the application of ``quasiclassical approximation'' in Ref.\cite{wang07} which treats system classically neglecting all quantum fluctuations and random noises from the baths as quantum mechanically correlated. At high temperatures, where thermal fluctuations predominate over quantum fluctuations, the system is inherently classical. In the opposite limit, the strength of the anharmonicity in the quartic onsite model is weaker if the temperature is lower. Here the anharmonicity can be treated perturbatively in an effectively harmonic system. So in these two limits the so called ``quasiclassical approximation'' is valid. But in the intermediate temperature where the anharmonicity has significant strength, quantum fluctuations due to non-commutativity of the operators play much important role in lower dimensions. Then the ``quasiclassical approximation'' ceases to be correct. Thus, though use of the ``quasiclassical approximation'' looks attractive, it has probably little application for real problems of quantum transport where phonon-phonon interaction is crucial. Finally one main feature of our analysis is that the effective transport mean free path distinguishing ballistic regime from diffusive one, emerges naturally in the study.

\section{acknowledgments}

The author would  like to express his thanks to  Abhishek Dhar for many useful discussions and Abhishek Chaudhuri for a critical reading of the manuscript.


\end{document}